\documentstyle[12pt, aaspp4, psfig]{article}

\def\refer#1{\noindent\hangindent=24.0pt\hangafter=1{#1}\par}

\def\gs{{_>\atop^{\sim}}}
\begin{document}
\centerline{\Large\bf A Structure for Quasars}
\bigskip

\centerline{Martin Elvis}
\centerline{Harvard-Smithsonian Center for Astrophysics, 
Cambridge MA 02138, USA}

\centerline{\tt version: 3pm, 18 June 2000}
%\authoremail{elvis@cfa.harvard.edu}
%
%%%%%%%%%%%%%%%%%%%%%%%%
\begin{abstract}
This paper proposes a simple, empirically derived, unifying
structure for the inner regions of quasars. This structure is
constructed to explain the broad absorption line (BAL) regions,
the narrow `associated' ultraviolet and X-ray `ionized' absorbers
(NALs); and is also found to explain the broad emission line
regions (BELR), and several scattering features, including a
substantial fraction of the broad X-ray Iron-K emission line, and
the bi-conical extended narrow emission line region (ENLR)
structures seen on large kiloparsec scales in Seyfert images.

The model proposes that a funnel-shaped thin shell outflow
creates all of these features. The wind arises vertically from a
narrow range of radii on a disk at BELR velocities. Radiation
force then accelerates the flow radially, so that it bends
outward to a cone angle of $\sim$60$^{\circ}$, and has an
divergence angle of $\sim$6$^{\circ}$, to give a covering factor
of $\sim$10\%.  When the central continuum is viewed from the

side, through this wind, narrow high ionization `associated'
ultraviolet absorption lines and the X-ray `ionized absorbers'
are seen, as in many low luminosity active galactic nuclei. When
viewed end-on the full range of velocities is seen in absorption
with a large total column density, giving rise to the broad
absorption lines systems seen in a minority of quasars, the
BALQSOs.

The wind is both warm ($\sim 10^6~$K) and highly ionized. This
warm highly ionized medium (WHIM) has a density of $\sim
10^9~$~cm$^{-3}$, putting it in pressure equilibrium with the
BELR clouds; the BELR is then a cool phase embedded in the
overall outflow, avoiding cloud destruction through shear. The
wind has the correct ionization parameter and filling factor for
this. The high and low ionization zones of the BELR correspond to
the cylindrical and conical regions of the wind, since the former
is exposed to the full continuum, while the latter receives only
the continuum filtered by the former.

The warm wind is significantly Thomson thick along the radial
flow direction, producing the polarized optical continuum found
in BALs, but is only partially ionized, creating a broad
fluorescent 6.4~keV Fe-K emission line, and $>$10~keV Compton
hump.  The conical shell outflow can produce a bi-conical matter
bounded NELR.

Luminosity dependent changes in the structure, reducing the
cylindrical part of the flow, or increasing the mean angle to the
disk axis and decreasing the wind opening angle, may explain the
UV and X-ray Baldwin effects and the greater prevalence of
obscuration in low luminosity AGN.

\end{abstract}

%%%%%%%%%%%%%%%%%%%%%%%%
\section{Introduction}
\label{intro}

This paper proposes a simple unifying structure for the inner
regions of quasars. It is almost universally acknowledged that a
massive black hole lies at the core of all quasars, and it is
widely believed that this black hole is surrounded by an
accretion disk that emits the powerful continuum radiation
characteristic of quasars. Yet the slightly larger region around
the continuum source, which produces most of the prominent and
much studied emission and absorption features in quasar spectra,
is not well understood. It is this region that we address.

Any model of quasars and active galactic nuclei (AGNs) needs to
explain self-consistently a wide range of emission and absorption
line phenomena. At a minimum the pieces of this jigsaw include:
the $\sim$5000~km~s$^{-1}$ broad emission lines present in all
AGN; the $\sim$0.1$c$ ($\sim$30,000~km~s$^{-1}$) broad absorption
lines seen in some 10\% of quasars; and the highly ionized
$\sim$1000~km~s$^{-1}$ outflows seen in narrow absorption lines
in the ultraviolet and X-ray spectra of about half of Seyfert
galaxies.  This is a challenging array of observations, and they
lead to constraints which are hard to satisfy.  Each phenomenon
has been the object of many studies that have defined its
properties in detail over the last two or three decades. Still
missing though is a structural and dynamic context that fits
these apparently disjoint elements together (Peterson 1997;
Krolik 1999).  The most obvious geometries are a turbulent,
spherical distribution of clouds and a rotating disk with an
equatorial wind. While both structures have their successes, they
also have problems. One difficulty in particular stands out: how
can they explain the simultaneous presence of outflowing material
at 1000 km s$^{-1}$ and at 30,000 km s$^{-1}$ implied by the
absorption lines?

Beginning from this puzzle we propose the following,
phenomenologically based, geometric and kinematic structure: An
accelerating outflow with a funnel-shaped thin shell geometry is
responsible for all of these features. This is the simplest
geometry that can explain both the broad and narrow absorption
line observations. We find that having set up this structure it
also explains a wide variety of other emission line and
scattering phenomena. With minor extensions, this structure could
also explain a number of luminosity related effects.

The scope of this paper is to outline how this structural model
can link these diverse phenomena empirically.  In individual
areas, preceding studies have often argued for the same or
related conditions.  Fitting all of these areas together has not
been achieved before, and it is the structure proposed here that
makes this possible.  Many of the details and theoretical
consequences are not investigated here, but are explored
semi-quantitatively in the following sections to see if obvious
conflicts with observation arise.  At the end of the paper some
theoretical work is referred to that promises to give the model a
good physical basis.

\section{Overview}
\label{overview}

\begin{figure*}
\label{structure}
\centerline{ 
\psfig{figure=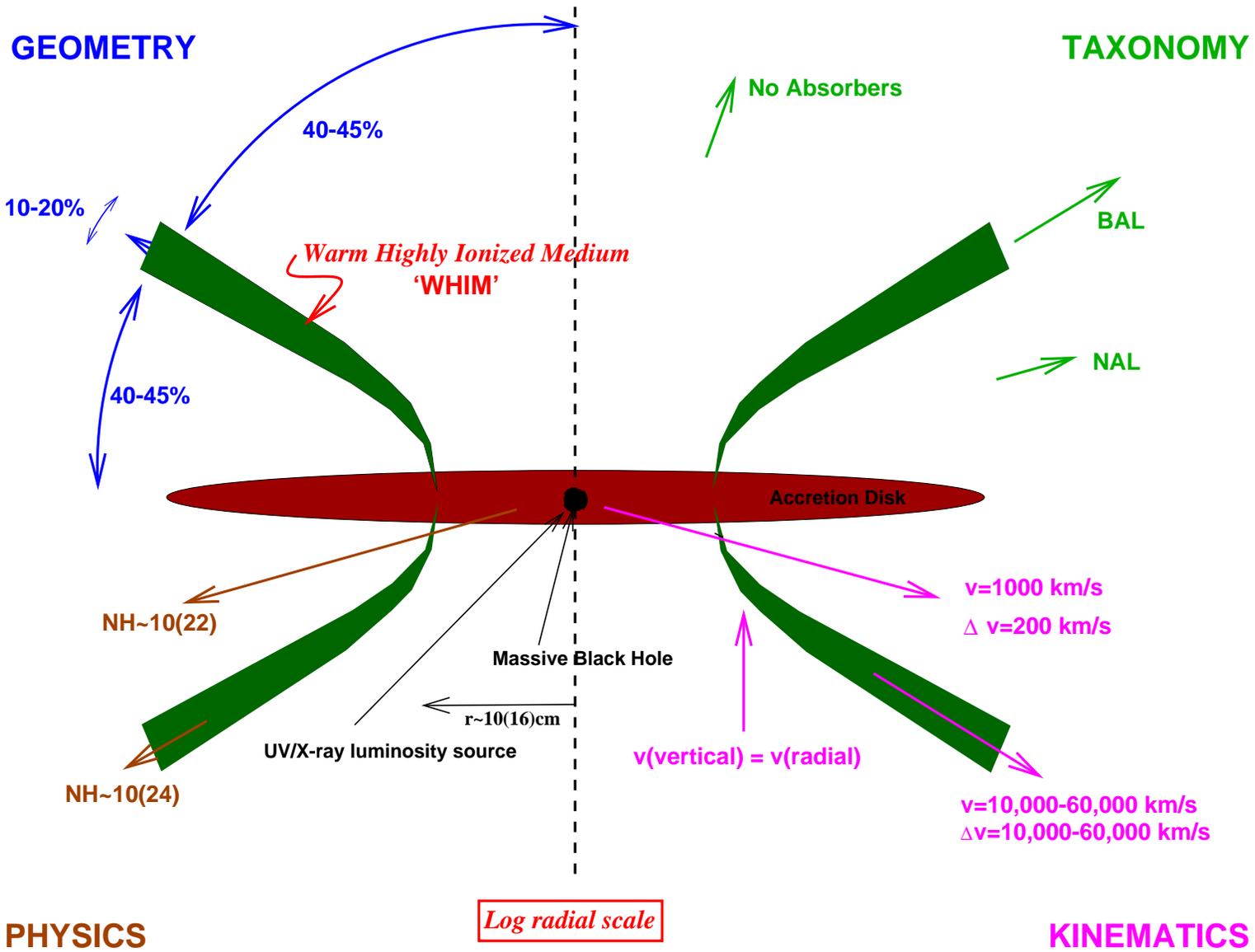,width=6in}
}
\caption{The proposed structure. The four symmetric quadrants
illustrate (clockwise from top left): the opening angles of the
structure; the spectroscopic appearance to a distant observer at
various angles; the outflow velocities along different lines of
sight; some representative radii (appropriate for the Seyfert 1
galaxy NGC~5548) and some typical column densities.
}
\end{figure*}

In outline\footnote{For brevity all references are deferred to
the following detailed discussion.}, a flow of warm gas first
rises vertically from a small range of radii on an accretion disk
rotating, tornado-like, with the initial Keplerian disk velocity,
comparable with BEL velocities. The flow then angles outward, and
accelerates to BAL velocities, until it makes an angle of
$\sim$60$^{\circ}$ to the quasar axis, with a divergence angle of
$\sim$6$^{\circ}$--$\sim$12$^{\circ}$ (figure~\ref{structure}).
Viewed along the flow a BAL is seen.  Dust in low luminosity
objects prevents the BALs being seen in Seyfert~1 galaxies. (They
appear as Seyfert~2s.)

Viewed across the flow NALs and X-ray warm absorbers are
observed.  Viewed from above no absorbers are apparent.  The
angles are set in order to produce the correct ratios of Narrow
Absorption Line (NAL), Broad Absorption Line (BAL) and
non-absorbed quasars.  The medium is warm ($\sim$10$^6$~K), has a
high density ($n_e\sim 10^{9}~cm^{-3}$), and is highly ionized,
as required by X-ray observations. We call this the Warm Highly
Ionized Medium (WHIM).

These properties make the WHIM the confining medium for the
clouds of the broad emission line region (BELR), for which it has
the correct pressure, ionization parameter, radius and filling
factor. If the wind origin radii span a factor 2, with decreasing
density at larger radii then the two zones of the BELR have a
natural origin: the high ionization BELs originate primarily in
the inner region where the clouds are exposed to the full
ionizing continuum, while the low ionization BELs originate in
the outer region where they are exposed only to the continuum
filtered through the WHIM. An outflowing wind with embedded
cooler clouds suffers none of the shear stress problems of fast
moving clouds in a stationary atmosphere that have seemed to make
pressure confined BELR clouds implausible, and the thin shell
geometry avoids the Compton depth problem previously encountered
in such models.  The WHIM also produces the high ionization
`coronal' optical emission lines.

Along the conical outflow direction the WHIM has significant
optical depth to electron scattering ($\tau\sim$1), as required
by spectropolarimetric observations of BALs and suggested by
X-ray spectra. This allows the WHIM to be the source of some, and
perhaps all, of the five scattering phenomena in AGN.  Scattering
off the far side of the flow explains (1) the $\sim$10\%
polarized continuum seen in BAL troughs, and the 0.5\%
polarization of non-BAL quasars, and (2) the strong continuum
polarization in the UV. In addition (3) the `mirror' seen in the
20\% of Seyfert~2 galaxies with polarized broad emission lines
can have the same origin if, in low luminosity AGN, the flow is
dusty. X-ray reflection features, (4) the `Compton hump' above
10~keV and (5) the fluorescent Fe-K line at 6.4~keV will arise
from continuum reprocessing in the warm conical shell wind, and
will provide, at a minimum, a significant fraction of the
observed features.

The model is summarized in figure~\ref{structure}.  The top left
quadrant of figure~\ref{structure} (`geometry') shows the
required angles. The top right quadrant of figure~\ref{structure}
(`taxonomy') shows which lines of sight give rise to which type
of absorber, while the lower right quadrant (`kinematics') shows
typical velocities. The lower left quadrant (`physics') shows the
relevant column densities and optical depths.

This geometry arises naturally if a disk instability creates a
wind. For example, radiation pressure dominates a disk inside a
critical radius. At smaller radii something (e.g. a hot corona)
must suppress the wind, leaving only a narrow boundary region on
the disk from which the wind can escape. Centrifugal action and
radiation pressure then bends and accelerates the flow.  At large
radii this geometry may produce the bi-conical narrow emision
line structures.

High luminosity quasars differ from the lower luminosity AGN in
several features: the lower CIV EW (Baldwin effect), the rarity
of NALs, and the weakness of X-ray scattering features in high
luminosity quasars. Small changes in the outflow shape
(specifically in the outflow opening angle, divergence angle and
the height of the cylindrical region), may explain these. Such
shape changes could result from increased radiation or cosmic ray
pressure at high luminosities.

%%%%%%
%%%%%%

%%%%%%%%%%%%%%%%%%%%%%%%%%%%%%%%%%%%%
\section{Presentation of the Quasar Structure}
\label{presentation}

%%%%%%%%%%%%%%%%%%%%%%%%%%%%%%%%%%%%%
\subsection{A Conical NAL Geometry}
\label{cones}

\begin{figure*}
\label{startingpoint}
\centerline{ 
\psfig{figure=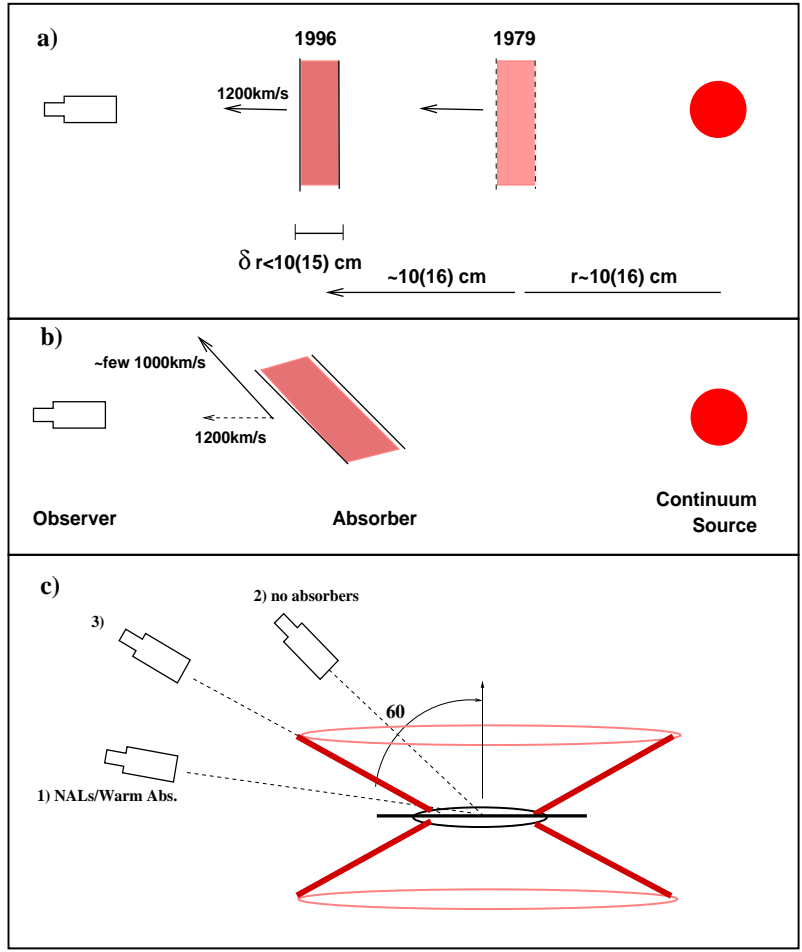,height=16cm,width=12cm,angle=0}
}
\caption{Geometry of the NAL/X-ray warm absorber: (a) Apparent
outflow rate would double the continuum-absorber separation in 20
years, incompatible with observations.  (b) A steady state flow
crossing our line-of-sight reconciles the observations. (c) The
simplest nuclear structure compatible with (b) is a bi-cone.}
\end{figure*}

The intensely studied active galactic nucleus (AGN) NGC~5548 has
had a NAL (of column density of 4$\times$10$^{21}$cm$^{-2}$) with
a outflow velocity of 1200~km~s$^{-1}$, for over 20~years without
significantly altering its high ionization state (Shull and Sachs
1993; Mathur, Elvis and Wilkes 1999).  Yet a radial flow would
have at least doubled its distance from the ionizing continuum
source over this time (figure~\ref{startingpoint}a; Mathur,
Elvis, and Wilkes 1995), which would increase the column density
in the trace CIV ion by a factor of $\sim$100, contrary to
observations.  Continuous flow along the line of sight is ruled
out, since the absorber is constrained to be thin in that
direction [thickness $<10^{15}$cm, c.f. distance from continuum
$=2\times 10^{15}-2\times 10^{18}$cm, if the UV NAL and the high
ionization (dominated by OVII, OVIII) X-ray absorption both arise
in the same material (Mathur, Elvis, and Wilkes 1995)
\footnote{
Statistically the UV absorption features and the X-ray warm
absorbers are closely related (Crenshaw et al. 1999), and can
even be used to predict one another's presence (Mathur, Wilkes \&
Elvis 1998). The first {\em Chandra} absorption line results
(Kaastra et al. 2000, Kaspi et al. 2000) show that the high
ionization X-ray absorption lines also have outflow velocities
matched to those of the UV absorbers. There remain difficulties
in fitting the observed UV and X-ray line strengths with a one
component model in some cases.
}
]. Instead a steady state flow {\em across} our line of sight,
with only a component of the velocity in our direction, provides
a natural explanation for this constancy of ionization
(figure~\ref{startingpoint}b; Mathur, Elvis and Wilkes 1995,
1999).

We now note that, given the above `continuous flow' requirement,
the simplest geometry for the NAL in NGC~5548 is a thin
bi-conical shell (figure~\ref{startingpoint}c) with a radial
outflow velocity that will be larger, possibly much larger, than
shown by the NAL. The thinness of the shell implies that the flow
arises at a limited range of radii on a disk.  The opening angle
of the conical shell is given by the ratio of NALs to absorption
free AGN (1:1) (Reynolds 1997; Crenshaw et al.~1999).  This ratio
implies that the flow makes an angle of 60$^{\circ}$ to the disk
axis, dividing the absorbed and unabsorbed solid angles equally
(figure~2c)

%%%%%%%%%%%%%%%
%%%%%%%%%%%%%%

%%%%%%%%%%%%%%%%%%%%%%%%%%%%%%%%%%%%%
\subsection{Broad and Narrow Absorption Lines: two views of the
same outflow}
\label{BALNAL}

A conical outflow has three distinct viewing directions
(figure~\ref{cones}c): above the cone, across the cone, and down
the length of the cone.  This third view will produce higher
velocity absorption in a small fraction of quasars. The flow, if
accelerated by radiation pressure must have a divergence angle,
since the illumination is not parallel. For an angle of
60$^{\circ}$ to the pole, the flow will have a height comparable
to its radius. The angle subtended to the accelerating continuum
will be given by the thickness of the flow, which is $\sim$0.1
radii, i.e. 6$^{\circ}$, which implies a covering factor of
$\sim$0.1. There is a population of quasars that shows these
characteristics, the Broad Absorption Line quasars (BAL quasars).

All radio-quiet quasars seem to contain high velocity outflows
($v\sim 0.1-0.2c$, 10-20 times larger than NALs) as shown by
studies of broad absorption lines (BALs) in quasars (Weymann et
al 1991; Hamann, Korista, and Morris 1993).  Any model of quasar
structure must then include BAL outflows. BALs are observed
directly in 10-20\% of radio-quiet AGN and so cover a similar
fraction of the solid angle around the continuum source (Weymann
et al 1991; Hamann, Korista, and Morris 1993). The geometry of
BAL flows is not known, but may well have conical shell
geometries (Murray et al 1995; Ogle 1998).

We propose that NALs and BALs are two views of the same flows. In
NALs we look across the flow direction, while in BALs we look
down the length of the flow.  BALs had been thought to have
inadequate column densities ($\sim 10^{20}$-10$^{21}$cm$^{-2}$),
similar to NALs, and relatively low ionization (Turnshek 1988).
Now however X-ray results on broad absorption line quasars
(BALQSOs) (Mathur, Elvis, and Singh 1996; Gallagher et al 1999,
Mathur et al. 2000) strongly suggest absorbing column densities
$\gs few\times 10^{23}~$cm$^{-2}$, 2-3 orders of magnitude larger
than expected from optical data alone. The relatively weak
optical absorption then requires the BALs to have high ionization
parameters, comparable with those of NALs (Hamman 1998).  The
detection of NeVIII, OVI and SiXII in some BALs (Telfer et al
1998) supports high BAL ionization parameters.  

Variability studies give some support for the existence of large
structures in quasars on this size scale: microlensing events in
the double quasar Q0957+561 contain a low amplitude signal with
delay times of 52--224 days (Schild 1996), implying luminous,
possibly reflective, structure with a 10\% covering factor on a
size scale appropriate for being the BAL outflow (see \S
\ref{scatterer}); a continuum lag of 100 days is seen in NGC~3516
(Maoz, Edelson \& Netzer 1999) also implying a reflecting
structure on this scale.

A covering factor of 10\% for BALs implies a divergence angle of
6$^{\circ}$ (figure~\ref{structure}, top left quadrant,
`geometry'); 20\% implies a wider angle of 12.2$^{\circ}$.
Continuity implies a decreasing density in the wind with radius
which will maintain a constant ionization parameter, apart from
the attenuation of the ionizing radiation by absorption and
scattering.

\begin{figure*}
\label{funnel}
\centerline{
\psfig{figure=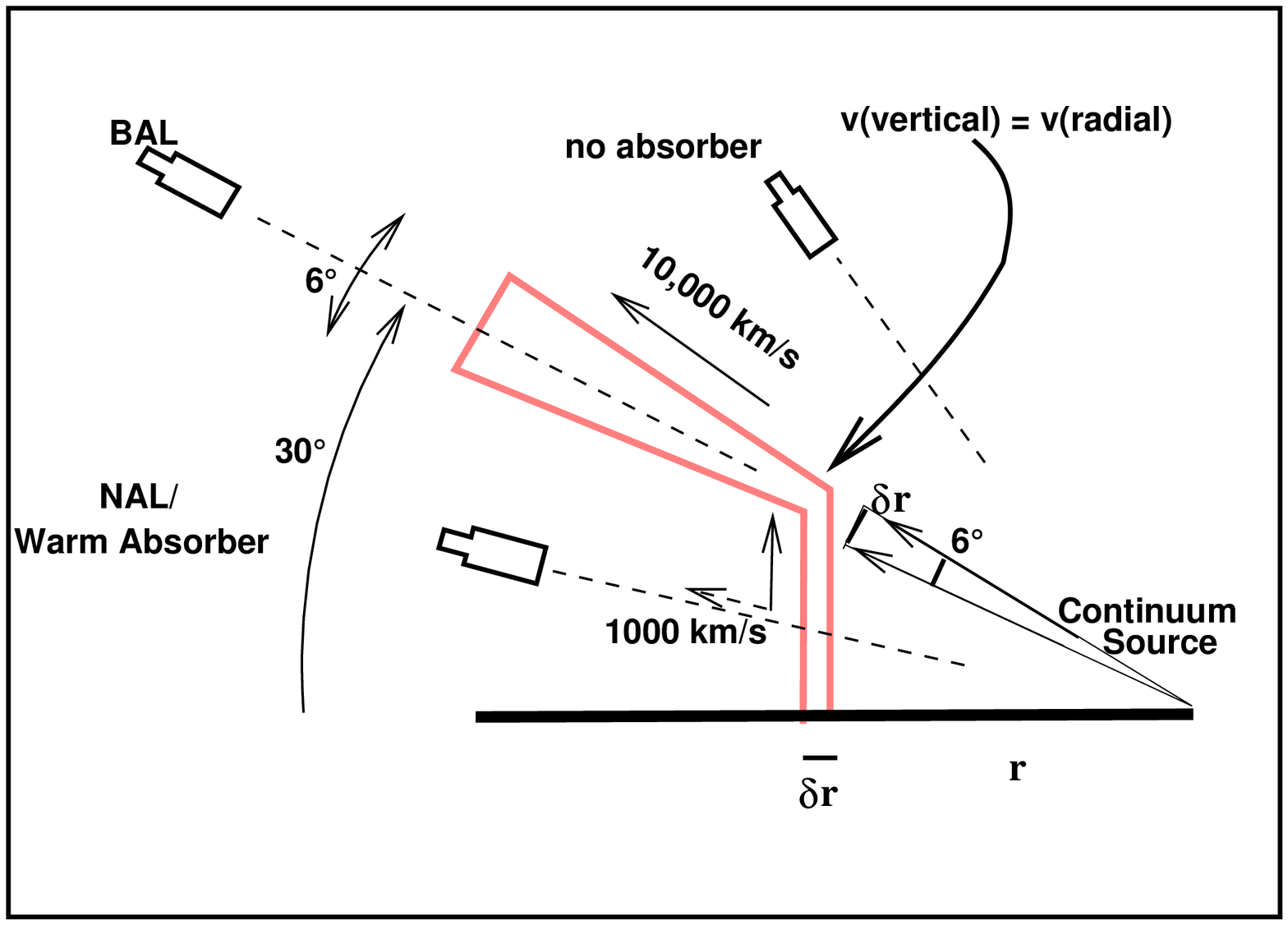,width=5in,angle=0}
}
\caption{A `funnel' is the simplest geometry that allows NALs and
BALs to arise in the same outflow.  }
\end{figure*}

A simple bi-cone though, is not compatible with the data.  The
velocity of the NALs, $v(NAL)\sim 1000~km~s^{-1}$ is only
$\sim$1/20 that of a BAL, implying that the typical angle through
the flow at which associated absorbers are found is 87$^{\circ}$,
inconsistent with the angles derived above. The simplest geometry
that removes this problem is to have the flow begin as a locally
vertical flow, which is appealing on the grounds of symmetry.
Most NALs then arise in a quasi-cylindrical region, and would be
viewed almost directly across the flow (figure~\ref{funnel}). The
cylindrical flow is viewed in NALs between 90$^{\circ}$ and
57$^{\circ}$ (figure~\ref{funnel}) for a mean angle of
80$^{\circ}$, so the vertical outflow velocity in this region is
typically some 6~times greater than the observed NAL blueshift,
i.e. comparable to the BELR velocities (FWHM).  A cylindrical
flow will inevitably curve outward at some point as vertically
displaced gas elements will have excess kinetic energy for their
increased radial distance from the central mass (deKool 1997).
The whole flow then has a funnel-shape.

%%%%%%%%%%%%%%
%%%%%%%%%%%%

This shape has a natural physical interpretation.  Radiation or
cosmic ray pressure can cause the flow to bend radially outward
and accelerate it to the large BAL velocities (deKool 1997). In
this picture the onset velocities of BALs relative to the
emission line peaks, occur at the point where the flow turns
outward to become part of the BAL line-of-sight. This will happen
when the radial velocity becomes comparable with the vertical
velocity (figure~\ref{funnel}).  The range of vertical velocities
will be similar to the BAL `detachment velocities',
$\sim$0-5000~km~s$^{-1}$ (Turnshek 1988; Weymann et al 1991).
N. Murray (2000, private communication) suggests that Low
ionization BALS, which have narrower widths than high ionization
BALS, might arise from a shielded low ionization region just
outside the high column density X-ray `warm absorber' producing
part of the outflow, extending for about a factor 2 in radius.
This arrangement of matter proves to be valuable in understanding
BELR structure too (see \ref{BELR}).

%%%%%%%%%%%%%%%%%%%%%%%%%%%%%%%%%%%%%%%%%%%%%%%%%%
\section{Implications of the Structure}
\label{implications}

Given the structure just presented, other parts of the AGN puzzle
fit quite naturally.  Several of the solutions given below have
been proposed individually before, and references are given, but
they have not before all been fitted together in a single scheme,
as here.  The structure presented here allows the
solution. 

\begin{figure*}
\label{3D}
\centerline{\psfig{figure=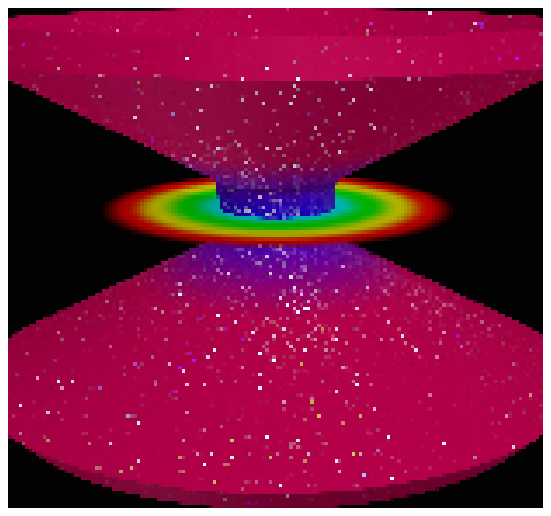,height=6cm,width=12cm,angle=0}}
\vskip 0.1cm
\centerline{\psfig{figure=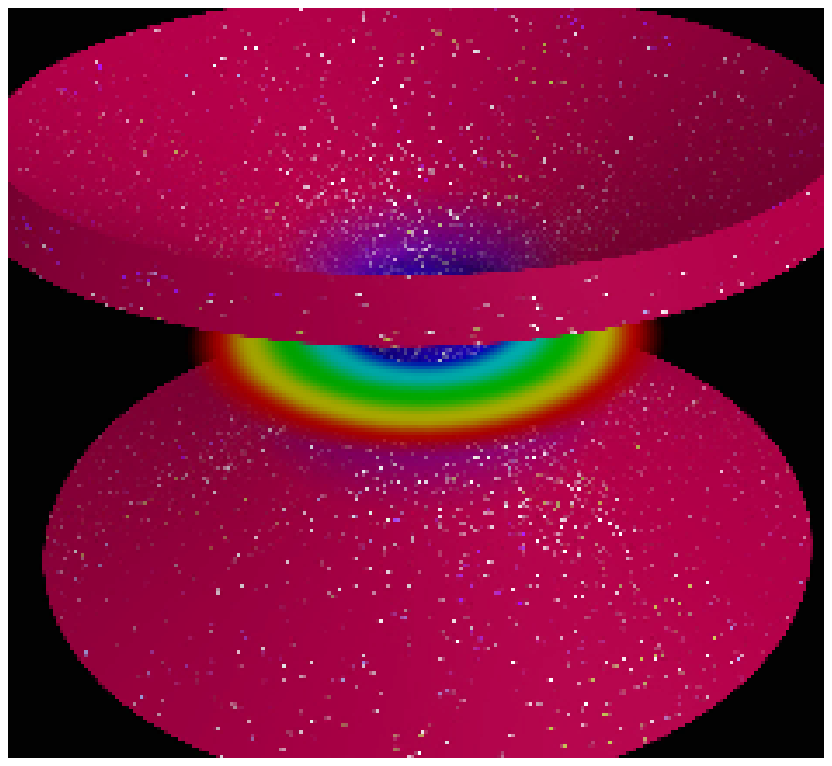,height=6cm,width=12cm,angle=0}}
\centerline{\psfig{figure=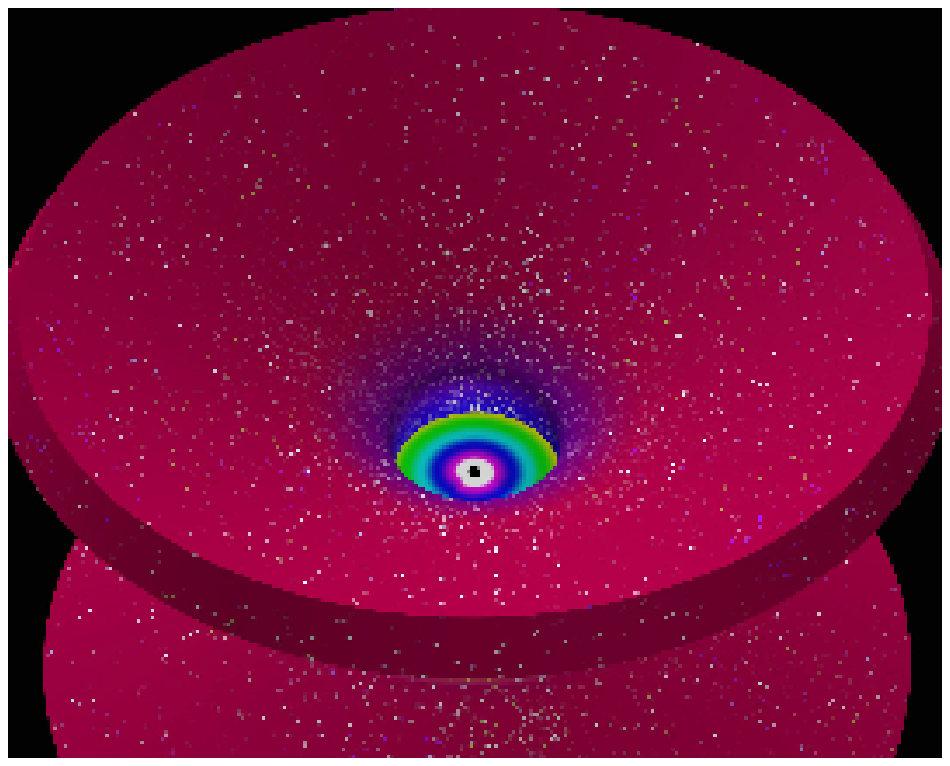,height=6cm,width=12cm,angle=0}}
\caption{3-dimensional views of the proposed structure: {\em top}
from the side, through the NAL producing outflow; {\em middle}
down the length of the accelerating, BAL-producing outflow toward
the continuum source (note the visibility of a large area of the
far side of the flow); {\em bottom} over the top with an
unobscured view of the continuum source. The multi-colored disk
represents the accretion disk. The small white dots represent
BELR clouds embedded in the flow. (The sharp edges are an
artifact of the 3-D rendering program.)}
\end{figure*}

Figure~\ref{3D} shows a 3-dimensional rendering of the
structure. The multi-colored disk represents the accretion
disk. The top figure show the side view, viewing the 
continuum source through the cylindrical part of the flow, where
absorption produces the NAL; the middle figure shows the view
directly down the conical flow, where the accelerating flow
produces a BAL spectrum; the lower figure shows the view over the
top of the flow, where there is a clear line of sight to the
continuum source. Embedded in the flow are small condensations
which are the BEL clouds. We discuss these next.

%%%%%%%%%%%%%%%
%%%%%%%%%%%%%%%

%%%%%%%%%%%%%%%%%%%%%%%%%%%%%%%%%%%%%%%%%%%%%%%%%%
%\subsection{The WHIM is the Coronal Line Region}
%\label{coronal}
%
%Many AGNs show high ionization forbidden `coronal' lines,
%particularly of iron ([FeX]6375\AA, [FeXI]7892\AA,
%[FeXIV]5303\AA). It is natural to investigate whether these are
%related to the ionized X-ray absorber. Porquet et al (1999)
%compared the conditions in the coronal lines with those in
%ionized absorbers and found reasonable agreement. They
%encountered a difficulty however with assuming a common
%origin. Since half of all AGN show ionized absorbers
%(\S\ref{cones}) they reasonably assumed a covering factor,
%$f_C$=0.5 for the region producing the emission lines. The
%resulting model overproduced the coronal lines and required
%densities as high as 10$^{10}$~cm$^{-2}$.  However $f_C$=0.5
%maximises the emission line equivalent widths (Netzer 1993). The
%BAL flow, which contains most of the WHIM gas, instead has
%$f_C$=0.1, reducing the predicted line strengths by a factor 5.
%There is then no conflict in having the WHIM produce the coronal
%emission lines. 
%
%%%%%%%%%%%%%%%%%%%%%%%%%%%%%%%%%%%%%%%%%%%%%%%%%%
\subsection{WHIM and BELR: two phases of the same
Medium}
\label{BELR}

The broad absorption line and broad emission line regions are
closely related. Lee \& Turnshek (1995) see a correlation of
their line widths.  Also the size of the high ionization
BELR (in CIV) in NGC~5548 is 10~light-days (Peterson 1997),
$\sim$10$^{16}$cm, consistent with the distance of the WHIM from
the ionizing continuum (\S\ref{cones}) (Mathur, Elvis, and Wilkes
1995). 

Conditions in the WHIM and BELR are also suggestively related: A
strong case can be made that the WHIM is sufficiently warm and
dense that it can pressure confine the BELR clouds, as proposed
here. Pressure confined BELR models have a long history (Matthews
1974) but are not currently favored (Peterson 1997). Nevertheless
the case for high ionization material in the vicinity of the BELR
has been well argued by Shields, Ferland \& Peterson (1995) and
by Hamann et al (1995a,b) who detected broad NeVIII emission
lines.  That the NAL medium confines the BELR clouds has been
suggested before (Turner et al 1993; Kaastra, Roos and Mewe 1995;
Marshall et al 1997).  Additional evidence for high pressure,
combined with the geometry proposed here, now make the case
stronger.

The evidence for the WHIM having a high pressure comes from X-ray
absorption and emission features.  X-ray absorption edges in some
AGN fail to respond immediately and linearly to changes in the
ionizing continuum as expected in equilibrium photoionization
models (Fabian et al 1994; McHardy et al 1995).  Both
non-equilibrium models and warm absorbers at $\sim$10$^6$K (where
collisional ionization competes with photoionization) can
reproduce these changes, and both require quite high densities,
$10^6<n_e<10^8$ (Nicastro et al 1999).  Higher densities are
needed in AGN where no delays are seen, (e.g. NGC~5548; Nicastro
et al 2000).

Strong high ionization OVII emission lines are seen in soft
X-rays in the Seyfert~1 galaxies NGC~5548 and NGC~3783 (George et
al. 1995, Nicastro et al. 2000, Kaspi et al., 2000, Kaastra et
al., 2000). However, a spherical geometry would produce EUV and
X-ray emission lines that were much too strong. Kaastra et
al. (1995) find a filling factor for the WHIM of $\sim$0.5\%, in
reasonable agreement with the 1\% of a conical shell with a
6$^{\circ}$ divergence angle.  To explain these lines requires a
thermal ionization component at T$\sim$10$^6$~K and $n_e\sim
10^9$cm$^{-3}$ (George, Turner, and Netzer 1995; Nicastro et al
1999).  This thermal gas has the correct ionization state and
column density to produce the X-ray absorbers and the UV NALs
(Kaastra, Roos, and Mewe 1995)
\footnote{ The EUV NeVII/NeVIII blend, SiVII emission lines
reported in NGC~5548 by Kaastra, Roos \& Mewe (1995) have been
disputed by Marshall et al. (1997). However, the conditions they
derive are sufficiently close to those required from the X-ray
OVII and variability data, that their other conclusions apply. We
predict then that the lines they report will be found with
observations of slightly greater sensitivity.}.
The pressure in the WHIM of the ionized absorbers in the NGC~5548
is $P_{WHIM}=10^{15}$cm$^{-3}$K, which is comparable to the
pressure in the BELR (Osterbrock 1989; Ferland et al 1992),
$P_{BELR, CIV}=10^{15}$cm$^{-3}$K. The factor $\sim$100 higher
density in the BELR requires that the ionization parameters of
the WHIM, $U_{WHIM}$ be a factor $\sim$100, higher than that of
the BELR, $U_{BELR}$.  This is close to the observed factor of 50
($U_{BELR, CIV}$=0.04 Peterson 1993; $U_{WHIM}$=2; Mathur, Elvis
and Wilkes 1995).  This may well be generally true, since NALs
(Crenshaw et al. 1999) have quite uniform $U$ as do BELRs
(Osterbrock 1989).  Kaastra et al. (1995) find (independent of
their EUVE data) that for the observed NGC~5548 continuum a
2-phase medium forms at the distance of the BELR, with
T=5$\times$10$^4$~K and T=$few \times$10$^5$~K, a slightly
narrower range than above, but in good qualitative agreement,
given the poorly known continuum from 100\AA -- 912\AA.

\begin{figure*}
\label{BEL}
\centerline{ 
\psfig{figure=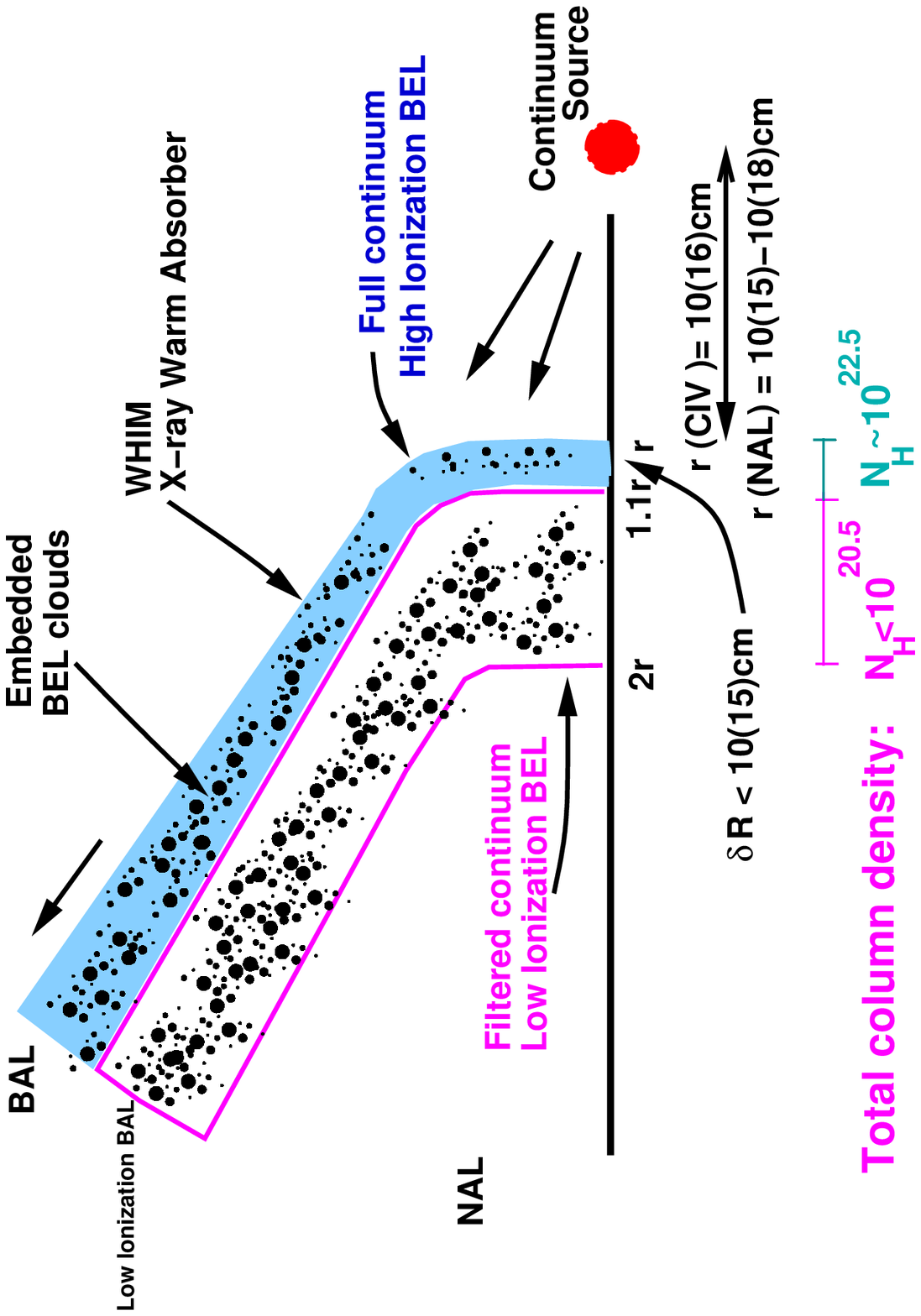,height=27cm,width=15cm,angle=-90}
}
\caption{The Broad Emission Line Region as a cool phase in the
Warm Highly Ionized Medium (WHIM) outflow. Distance scales are
shown for NGC~5548. The high ionization BELR is coincident with
the X-ray absorbing WHIM, while the low ionization BELR lies at
larger radii where it is shielded, by the WHIM, from the full
ionizing continuum.}
\end{figure*}

If the outflow spans a factor 2 in radius then the two BELR zones
deduced from observation (Collin-Souffrin et al. 1988, Goad et
al.  1999) are created naturally (N.  Murray 2000, private
communication).  These zones are: the `high' ionization zone
(e.g.  CIV) and the `low' ionization zone (e.g.  MgII)
(figure~\ref{BEL}, which has large optical depth and is mostly
heated and ionized by hard X-rays (Collin-Souffrin et al 1988).
The WHIM filters the ionizing continuum, preventing soft X-ray
flux (Ferland et al 1990) reaching the outer radii. This is
appealing since the low ionization lines have somewhat narrower
widths (FWHM) than high ionization lines, which would arise from
the $\surd 2$ lower Keplerian disk velocities at double the
radius, and is consistent with the line width vs.  inverse
square-root of the lag time relation found for several Seyfert
galaxies (Peterson \& Wandel 1999, 2000). A zone of about the
correct width was noted by Nicastro (2000) in his model which
relates BEL width to Keplerian disk velocity which uses the disk
instability model of Witt, Czerny, \& \.{Z}ycki (1997) (see \S
\ref{models}).

%%%%%%%%%%%%%%%
%%%%%%%%%%%%%%

How can a factor $\sim$2 in radius be reconciled with the thin
shell required by the X-ray observations? The thin shell applies
to the high ionization `warm absorber' zone. A lower density
spanning a wider range of radii is allowed so long as the total
column density through the low ionization region is small enough
not to have significant optical depth in soft X-rays, notably in
oxygen, i.e. N$_H\le$10$^{20.5}$cm$^{-2}$. Since the warm
absorbers have N$_H\sim$10$^{22}$cm$^{-2}$ a strong radial
density gradient in the outflow is required. In the model of
Witt, Czerny, \& \.{Z}ycki (1997) a gradient will be present, but
may not be sufficiently steep (B. Czerny, 2000, private
communication).  Evidence for cold absorption associated with
warm absorbers has been seen in several X-ray spectra (Komossa
1999), and has been modelled with dust absorption, though low
N$_H$ cold gas may also be allowed. If this is confirmed then the
steep density gradient requirement may be lessened.

Some low ionization line emission could arise along the conical
flow, several light-months away from the continuum source (for
NGC~5548), which also will be shielded from the full continuum by
the WHIM.  MgII in NGC~3516 has a reverberation time comparable
with these sizes (Goad \& Koraktar 1997). In this case though the
WHIM is being accelerated and so the lines should appear
broader. A low contrast `Very Broad Line Region' (VBLR) is
present as shown by variability (Ferland, Korista, and Peterson
1990) and polarimetry in some AGN (Goodrich and Miller 1994;
Young et al 1999) with about double the width of the normal
(variable, unpolarized) BELR. These lines may arise in the
conical flow region.

Models in which the BELR clouds are confined in pressure
equilibrium in a 2-phase medium (Krolik, McKee and Tarter 1981)
have fallen out of favor due to lifetime problems (Matthews 1986)
and high Compton depth.  In the thin shell steady state outflow
picture presented here the geometry gives a medium which has a
low Compton depth except along the flow.  Moreover, the
individual clouds do not need long lifetimes, since they leave
the BEL emitting, vertical, part of the flow on an outflow
timescale, comparable to the dynamic timescale.  Moreover they
are not pulled apart by strong drag forces, since they co-move
with the confining medium.  Only a small energy input, $\sim
10^{-4}$ relative to the bulk kinetic energy of the outflow, is
needed to heat the NAL gas to 10$^6$~K.

Radiation driven flows commonly suffer instabilities and shocks
(e.g. Williams 2000) which may well heat the WHIM (e.g. Feldmeier
et al 1997), and may also produce the fine structure in
the absorption line velocity profiles seen in NALs (Mathur, Elvis
and Wilkes 1999; Crenshaw and Kraemer 1999) and BALs (Turnshek
1988).  In BALs the constancy of this velocity structure (Weymann
1997) suggests standing shocks, although in NALs fine structure
variability has been seen arguing for some transverse motions
(Crenshaw et al. 1999).  Shock heating along the flow may also
prevent the rapid adiabatic cooling of the WHIM.  If the WHIM
were in free expansion at the hot phase sound speed
(v(10$^6$K)$\sim$200~km~s$^{-1}$) and was moving away from the
continuum source at $v(outflow)\sim 20,000~$km~s$^{-1}$ in order
to make BALs, then an opening solid angle of only 1\% will
result. The flow must then be forced into supersonic lateral
expansion by some extra pressure, perhaps magnetic, to cover the
necessary 10\%--20\%.

If we provisionally accept the identification of the BELR as
condensations in the larger WHIM outflow, is this consistent with
the known properties of the BELs? Emission lines give weaker
constraints than absorption lines because the information they
carry has been integrated over a volume distribution of material,
rather than a line, convolving together ranges of density and
ionization state to produce the emission line. The prediction of
BEL profiles is thus subject to the uncertainties of many
parameter models. Nevertheless a funnel-shaped distribution for
the BELR seems to be consistent with the BEL profiles and widths.
The distribution of BEL widths can be reproduced well by
axisymmetric, but non-spherical, geometries (Rudge and Raine
1998), while the broad emission line profiles arise naturally in
a wind (Cassidy and Raine 1993; Murray and Chiang 1995).  The
model predicts orientation effects in BEL profiles. Edge-on AGN,
which will shown NALs, will be dominated by the disk Keplerian
velocities, while pole-on AGN, which will have no absorption
features, will be dominated by the outflow velocities, both
vertical and radial. Velocity resolved reverberation mapping of
the `edge-on' (because of the presence of NALs,
figure~\ref{structure}, top right quadrant) AGN NGC~5548 (Korista
et al 1995) shows that radial motion does not dominate for the
CIV line.  The cylindrical structure proposed here for the high
ionization BEL will be rotating with the disk, so rotation will
dominate in a NAL AGN, although a vertical velocity component
will also be present.  There is a tendency for high ionization
lines to be blueshifted relative to low ionization lines
(Peterson 1997; Wilkes 1984), which has no obvious interpretation
in this model.

The observation that the NALs absorb the BELs (Mathur, Elvis, and
Wilkes 1999; 1995) presents a difficulty, since it suggests that
the NAL is exterior to the BEL, rather than being
co-mingled. Several authors have noted though that this problem
is avoided if the BEL clouds primarily emit back toward the
continuum, and so are seen originating on the far side of the
flow (Hamann, Korista, and Morris 1993; Shields 1994; Shields,
Ferland, and Peterson 1995) and are seen through the near side.

%%%%%%%%%%%%%%%%%%%
\section{A Single Compton Thick Scatterer}
\label{scatterer}

Active Galactic Nuclei give ample evidence for scattering and
fluorescing media: (1) BAL quasars have 5\%-10\% polarized flux
filling in the BAL troughs (Cohen et al 1995; Goodrich and Miller
1995; Ogle 1998), requiring an ionized medium with significant
electron scattering optical depth ($\tau > 1$, $N_H>1.5\times
10^{24}$); (2) some type 2 Seyferts have highly polarized broad
lines (Antonucci and Miller 1985; Miller and Goodrich 1990) due
to electron scattering ($\tau > 0.1$) off a warm ($T \sim 3\times
10^5$K) medium (Miller, Goodrich, and Mathews 1991); (3) X-ray
spectra of Seyfert 1 galaxies often show an excess `hump' of
emission above 10~keV ascribed to Compton scattering ($\tau \gs
1$) (Piro, Yamauchi, and Matsuoka 1990; Pounds et al 1990;
Lightman and White 1988); (4) a broad fluorescent Fe-K line is
seen in many Seyfert~1 X-ray spectra peaking around 6.4~keV
(i.e. no more ionized than FeXVII), but extending down in a long
tail to $\sim$4--5~keV (Tanaka et al 1995; Nandra et al 1997b)
with implied Doppler velocities of up to $\sim 0.3~c$; (5)
rapidly rising UV continuum polarization toward short wavelengths
is seen in several quasars (Impey et al 1995; Koratkar et al
1995).

It seems unlikely that such an abundance of scattering phenomena
should all be independent, although {\em a priori} they could be,
and are usually modelled as such. The X-ray features are
normally, and plausibly, thought to arise in the inner regions of
an accretion disk, producing GR-distorted radiation (Bromley,
Miller \& Pariev 1998); the Seyfert~2 scatterer is believed to be
an extended, 1--50 parsec scale, cloud (Miller, Goodrich, and
Mathews 1991; Krolik and Kriss 1995); and the BAL scatterer is
assumed to be part of the BAL outflow (Goodrich and Miller 1995).
The UV continuum scatterer has been successfully modeled as a
mildly relativistic ouflowing wind from an accretion disk
(Beloborodov and Poutanen 1999), a situation strongly reminiscent
of the model proposed here.  The WHIM outflow has many of the
characteristics of all of these scatterers: high temperature and
ionization, a significant Compton depth, substantial covering
factor, and large velocities. The outflow must then contribute to
these features. Can it explain them entirely?

%%%%%%%%%%%
\subsection{BAL Scatterer}
\label{balscatterer}

In a BAL quasar, where we observe down the flow, the conical
shell outflow will produce an electron scattered continuum from
the far side of the outflow (figures~\ref{3D}c, \ref{scatterer})
which reaches a maximum 10\% polarization for a cone angle of
60$^{\circ}$ from the disk axis (Ogle 1998), the same angle we
find from other considerations (see \S 2). This component will
appear at all energies, including X-rays, where it will reduce
the apparent column density in existing X-ray spectra (Goodrich
1997b).  Viewed from any non-face-on angle above the flow, this
conical geometry will produce a slight net polarization that can
account (Ogle 1998) for the weak (0.5\% mean) optical
polarization of non-BAL radio-quiet quasars (Berriman et al.,
1990).  Spectropolarimetry already suggests that the BAL
scatterer is comparable in size to the BELR and may be cospatial
(Goodrich and Miller 1995; Ogle 1997, 1998; Schmidt and Hines
1999).

%%%%%%%%%%%%%%%
\begin{figure*}
\label{luminosity}
\centerline{ 
\psfig{figure=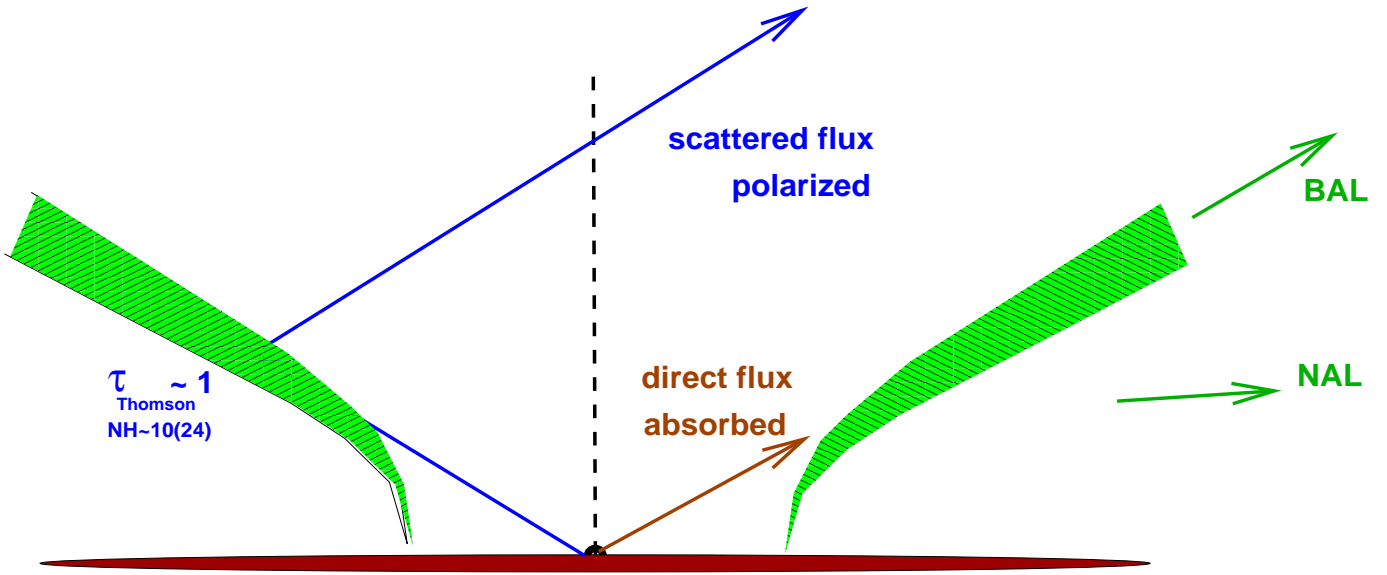,height=15cm,width=20cm,angle=-90}
}
\caption{Geometry of Thomson scattering off the flow. (see also
Ogle 1998.)}
\end{figure*}
%%%%%%%%%%%%%%%

%%%%%%%%%%%
\subsection{Seyfert 2 mirror}
\label{mirror}

The `mirror' in Seyfert 2 galaxies with polarized BELs has
comparable temperature, column density and outflow velocity
to the NALs (Beloborodov and Poutanen 1999; R.Antonucci 1999,
private communication).  If, like BALs, the `polarized BEL'
Seyfert~2s are viewed end-on through the flow then, as in BALs,
the polarized emission will be scattered off the far side of the
flow. 

Dust may be more abundant in low luminosity Seyferts than in
quasars (Edelson, Malkan and Rieke 1987), suggesting that Seyfert
BALs are not detectable because reddening removes the UV spectrum
where BALs are found, so making BALs rare in Seyferts, but
relatively common in (radio-quiet) quasars (Turnshek 1988).  [At
least some of the long sought `quasar~2s' (Almaini et al 1995;
Halpern and Moran 1998) are then the BAL quasars.]  About 20\% of
Seyfert~2s have polarized BELs (Kay 1994), however type~2
Seyferts are $\sim$4 times as common as type~1s (Huchra and Burg
1992; Rush, Malkan, and Spinoglio 1993; Comastri et al 1995) so
there are $\sim$80\% as many `polarized BEL' Seyfert~2s as all
Seyfert~1s, implying a covering factor close to 50\% (see
\S\ref{baldwin}).  The absence of large velocity shifts in the
polarized BELs of Seyfert 2s is not surprising since the opposite
side of the outflow is almost perpendicular to the line of sight.
Large transverse outflow velocities will produce a second order
Doppler shifts.  This could explain the 400 km s$^{-1}$ redshift
of polarized H$\beta$ in NGC~1068 (Miller, Goodrich, and Mathews
1991) if the transverse velocity is 15,000 km s$^{-1}$.
Spatially resolved Seyfert 2 mirrors (Capetti, Macchetto, and
Lattanzi 1997; Capetti et al 1996) on a $\gs$10 parsec scale are
probably too large to be a part of this structure, and have good
alternative explanations (e.g. the 70~pc scale, warped molecular
disk in NGC~1068, Schinnerer et al., 2000).

%%%%%%%%%%%
\subsection{X-ray Scatterer}
\label{xray}

Disk models for the X-ray broad Fe-K line and Compton hump have
difficulties in producing strong enough Fe-K emission
lines. Several theoretical explanations have been proposed
(stronger fluorescence from ionized plasma (Matt, Fabian, and
Ross 1991; \.{Z}ycki and Czerny 1994); enhanced abundances
(Turnshek et al. 1996, but see Hamann, Korista and Morris 1993,
Lee et al. 1999)
\footnote{In fact enhancements of $>$4 times solar are excluded
by the X-ray measurements.}
; anisotropic emission (George and Fabian 1991), but all have
difficulties.  The Seyfert galaxy IC4329A can have only a thin
(half opening angle $<$10$^{\circ}$) torus (Madejski et al.,
2000).  The inclusion of an additional Compton thick region, as
given by the proposed outflow, may solve this problem.  The
conical outflow has a 0.2$c$ velocity spread (defined by the BAL
widths), which is almost as large as the reported broad Fe-K
X-ray line widths, while also having the proper ionization state,
with Fe-XVII dominating (Nicastro et al 1999).  The 10\% -20\%
covering factor of the BALs is 2-5 times lower than those of disk
models and so would produce proportionately weaker Fe-K lines and
Compton humps. [Although, as noted above (\S 5.2), a covering
factor closer to 50\% may be more appropriate for the low
luminosity AGNs, in which broad Fe-K has been found.]  20\%-40\%
of the observed Fe-K and Compton hump features could then arise
in the outflow structure.

Outflow models for the broad Fe-K lines have been considered
previously: Fabian et al. (1995) noted that the strong observed
red asymmetry of the broad Fe-K lines could be produced from the
far side of the flow.  With a transverse column density N$_{\rm
H} \sim 10^{22}$ cm$^{-2}$, T$\sim 10^{6}$K and turbulent
velocities $<10^{3}$km s$^{-1}$ the WHIM is optically thick in
the FeXXV resonance line and so to locally emitted FeXXV
fluorescence.  Redshifted FeXXV instead falls in the continuum
where the WHIM is optically thin.  In this condition red
asymmetric lines may well be seen in NAL quasars, although
detailed calculation is needed to produce reliable line profiles.
However, Fabian et al. rejected outflow models both because of
their {\em ad hoc} nature, and because of an apparent conflict
with the smaller NAL velocities seen in the same objects. Our
model removes both objections (\S\ref{BALNAL}): the outflow is
already required to explain other quite different, absorption
line, phenomena; and our geometry reconciles the low absorption
line velocity across the flow, with the large emission line
velocity width integrated along the flow.

While the size of the WHIM outflow permits variations of the
broad Fe-K line over months, Fe-K line variability by a large
factor on a timescale of days would limit the contribution from
the extended conical shell outflow. Some broad Fe-K lines may
show such variations (Iwasawa et al 1999; Wang et al 1999), while
others do not (Georgantopoulos et al. 1999, Chiang et al. 2000),
including the Seyfert galaxy NGC~7314 which shows apparent 1-day
variability at low Fe-K velocities, but not at high velocities
(Yaqoob et al 1996). IC4329A was studied simultaneously with ASCA
and RXTE, so covering the Fe-K line and Compton hump with good
S/N, making it a particularly good example (Done et al.,
2000). In this observation the Fe-K line is clearly broad but,
with a width of $\sim$23,000~km~s$^{-1}$, it is distinctly
narrower than a relativistic disk fit. The line also does not
vary when the continuum changes by a factor 2. In a disk-torus
scenario this behavior is puzzling.  If a substantial fraction of
the broad Fe-K line arises from the BAL outflow then the lack of
response of the Fe-K line to continuum changes in these cases can
be understood.

%%%%%%%%%%%%%%%%%%%%%%%%%%%%%%%%%%%%%%%%%%%%%
\section{Discussion}
\label{discussion}

%%%%%%%%%%%%%%%%%%%%%%%%%%%%%%%%%%%%%%%%%%%%%
\subsection{Building BAL Column Densities}
\label{BALNH}

The expanding flow of figure~\ref{structure} naturally creates a
larger column density along the flown than across it. Integrating
the column density with mass conservation gives a factor increase
of $1/\theta$ radians, where $\theta$ is the flow divergence
angle.  (The 1/$r^2$ decrease in density renders the
contributions to column density from the outer parts of the flow
negligable.)  For $\theta$=6$^{\circ}$ this gives a factor
$\sim$10, i.e. 4$\times$10$^{23}$atoms~cm$^{-2}$ for
NGC~5548. This is somewhat less than a Thomson depth,
1.5$\times$10$^{24}$atoms~cm$^{-2}$, implying $\tau_{es}$=0.25
and an scattering of 23\% of the incident flux. This is somewhat
low to create the observed scattering effects, and more detailed
calculation is needed to test whether a larger column density is
required.

A higher column density outflow could be created by adding mass
to the WHIM. The BEL clouds offer a way of doing this. Once they
enter the radial part of the flow the BEL clouds become shielded
by a larger WHIM column density, so making the continuum they see
weaker in soft X-rays. Nicastro (1995) has looked at the effect
of steepening the ionizing ultraviolet to X-ray slope
($\alpha_{OX}$) on the 2-phase medium, and finds that the
instability disappears below some critical $\alpha_{OX}$. If this
applies in the conical part of the flow then the BEL clouds will
disperse, increasing the total column density.
In order to increase the column density by a factor 5-10 the mass
in the BEL clouds has to be 5-10 times greater than in the
WHIM. As yet there is no theory for the amount of matter that
goes into each phase of a multi-phase medium such as the quasar
outflow or the galaxy interstellar medium, so that a factor of
this size is acceptable. BEL clouds need a minimum column density
of 10$^{22}$cm$^{-2}$ (Krolik 1999), but there is no upper limit
and large values are sometimes invoked (Ferland et al., 1992).
This allows large values of M(BEL clouds)/M(WHIM), depending on
the size and number of the clouds. Using the same simplifying
assumptions as Peterson (1997), and a cylinder of thickness 1/10
of its radius for the vertical part of the flow, gives:
$\frac{r(BEL cloud)}{r(cylinder)}= 66 N_{cloud}^{-\frac{1}/{3}}$.
So for $N_{cloud}=10^6$, $\frac{r(BEL cloud)}{r(cylinder)}$=0.04,
and $r(BEL cloud)=4\times 10^{14}$cm, roughly 10 Schwarzchild
radii ($r_s$) for a 10$^8 M_{\odot}$ black hole, as deduced to be
present in NGC~5548 from reverberation mapping (Ho 1998). For
10$^9$ clouds these dimensions become $\frac{r(BEL
cloud)}{r(cylinder)}$=0.004, $r(BEL cloud)=4\times 10^{14}$cm
$\sim 1 r_s$. Smooth BEL profiles favor $N_c > 10^7$ (Arav et
al. 1998). These then are plausible numbers. 

Alternatively there might be a reservoir of more highly ionized
gas in the outflow. This is hinted at in the reports of an Fe-K
absorption edge in some Seyferts (Costatini et al. 2000, Vignali
et al., 2000).  In this gas all elements up to iron are fully
ionized.  Such a medium would add to the Thomson depth and Fe-K
fluorescence line strength without affecting the line absorption
in soft X-rays or the UV.

%%%%%%%%%%%%%%%%%%%%%%%%%%%%%%%%%%%%%%%%%%%%%
\subsection{Luminosity Dependences: the Baldwin Effect}
\label{baldwin}

High luminosity quasars differ from the lower luminosity Seyfert
galaxies in several ways: (1) High luminosity quasars have weaker
emission line equivalent widths, especially high ionization
emission lines (Osmer and Shields 1999; Espey and Andreadis
1999), than low luminosity quasars (the `Baldwin Effect'; Baldwin
1977). (2) In X-rays the weakness of Fe-K lines and Compton humps
in quasars is consistent with the Baldwin effect (Iwasawa and
Taniguchi 1993; Nandra et al 1997a). (3) NALs (in either UV or
X-rays) are rare in quasars, but common in the lower luminosity
Seyferts (Mathur, Elvis, and Wilkes 1999; Nicastro et al 1999).

We suggest that these three luminosity dependent effects are
directly related to changes in the funnel geometry
(figure~\ref{luminosity}).  The angles derived so far
(figure~\ref{luminosity}b) depend primarily on the statistics of
NALs among Seyfert~1 galaxies, which are AGNs of relatively
modest luminosity. If the quasi-vertical flow region is smaller
at high luminosities (perhaps because radiative acceleration is
stronger for high luminosity objects) then a smaller solid angle
will be exposed to the unattenuated continuum source, weakening
the BELR, particularly the high ionization lines.  There will
also then be a smaller solid angle from which the continuum can
be viewed through the WHIM, so NALs will be rarer.  If the cone
opening angle is also larger at low luminosities then larger
X-ray features will be produced, as noted above. A larger opening
angle is suggested by the greater number of `polarized BEL'
Seyfert~2s compared with BALs (\S\ref{scatterer}).

%%%%%%%%%%%%%%%
\begin{figure*}
\label{luminosity}
\centerline{ 
\psfig{figure=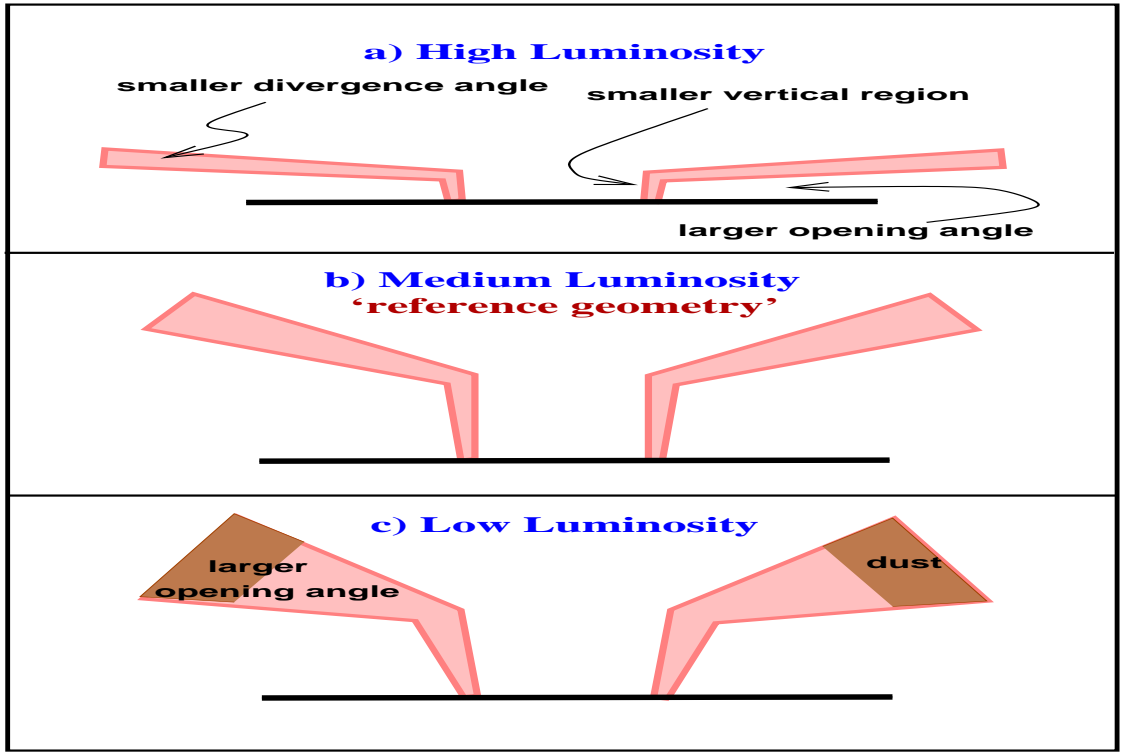,height=10cm,width=15cm,angle=-90}
}
\caption{Luminosity dependent changes in the outflow structure
that would account for observed luminosity correlations.}
\end{figure*}
%%%%%%%%%%%%%%%

%%%%%%%%%%%%%%%%%%%%%%%%%%%%%%%%%%%%%%%%%%%%%%%%%%%%%%%
\subsection{Biconical Structures and the Molecular Torus}
\label{torus}

For simplicity we have ignored the presence of accretion disk
flaring and of the molecular torus which is commonly invoked
(Antonucci and Miller 1985; Pier and Krolik 1992): (1) to explain
the presence of polarized broad line emission in otherwise narrow
lined (type 2) AGN (\S\ref{scatterer}); (2) to collimate the
ionizing radiation that leads to kpc-scale ionization cones of
the extended narrow line region (ENLR, Tadhunter and Tsvetanov
1989): and (3) to create the observed 4:1 ratio of obscured (type
2) to unobscured (type 1) AGN (\S\ref{scatterer}).  If we allow
the accretion disk or the torus to cover a large fraction of the
sky as seen from the continuum source then the angles in
figure~\ref{structure} will be raised toward the axis
substantially. However direct evidence for a nuclear torus is
weak. The maser source in NGC~4258 shows clearly that a thin
molecular disk is present in some AGN down to the 1~pc scale
(Greenhill et al. 1999), but does not show that a thick 2 $\pi$
covering torus is present.

Overall the need for a torus is weaker in this model: (1) We have
seen that the Seyfert~2s with polarized BEL could be viewed
through a dusty BAL (\S\ref{BALNAL}) and that other, larger
scale, asymmetric obscuring structures (e.g. the 70~pc-scale
warped molecular disk resolved in NGC~1068, Schinnerer et
al. 2000) can also produce polarized BEL spectra; (2) the conical
shell outflow could itself produce the ENLR bi-cone structures,
which would then be hollow matter bounded cones (as suggested by
Crenshaw \& Kraemer 2000 for NGC~5548), rather than
filled, ionization bounded structures; (3) the other, more
common, Seyfert~2s could be made in any of several ways: (a) the
large narrow emission line region NELR endures $\sim$1000~year
after the central source is extinguished (Lawrence \& Elvis
1982); (b) obscuration by large scale structures in the host
galaxy disks (Lawrence and Elvis 1982; Maiolino and Rieke 1995;
Simcoe et al 1998) or (c) in irregular non-nuclear obscuring
clouds (Malkan, Gorgian \& Tam 1998, REF 2000) can hide the small
BELR but not the much larger NELR in many AGN.

The structure presented here is not directly inimical to the
obscuring torus model, but a torus is less strongly required. The
WHIM outflow could in fact be considered as a form of the
obscuring torus.

%%%%%%%%%%%%%%%%%%%%%%%%%%%%%%%%%%%%%%%%%%%%%%%
\subsection{Discriminating between Wind Models}
\label{models}

Wind models for quasars have become widely discussed in recent
years (see reviews by deKool 1997, Vestergaard 2000).  In
particular the detailed models proposed by Murray, Chiang and
co-workers (Murray et al 1995; Murray and Chiang 1995, 1998), by
Cassidy \& Raine (1996) and, for binary systems, by Proga, Stone,
and Drew (1998, 1999) have many features in common with the
empirical picture developed here. This confluence is encouraging.
Murray et al. have a wind emerging from all disk radii and
accelerated into a wide cone by radiation pressure, which shows
BAL when viewed edge-on. The key geometric difference is that the
wind proposed here originates from a narrow range of radii and
rises almost vertically before bending outward into a cone. This
allows the NALs to be produced by viewing the flow edge-on.  The
main physical difference is that Murray et al. and Cassidy \&
Raine have a single phase medium. Murray et al. have different
ionization states being produced from a wide range of footprint
radii, producing a radially stratified BELR.  The flow proposed
here instead forms a 2-phase medium which originates in a narrow
footprint and the high and low ionization zones are created from
shadowing of the continuum by the WHIM.
%This geometry has at least a passing resemblence to that of a
%star in formation (specifically a T~Tauri star, Shu 19XX [???]).
Ionized absorbers are hard to produce from a radially stratified
BELR, unless the density gradient is large (\S \ref{BELR}), since
we will always be looking through low ionization material too,
which is X-ray opaque. However, the hydrodynamic models of Proga,
Stone and Drew (1998, 1999), Proga, Stone \& Kallman (2000) show,
under some conditions, `streamer' like high density structures
that resemble the structure proposed here.  Detailed calculation
may then show the models are identical.

A related class of models, hydromagnetic wind models, use a
magnetic field anchored in the disk to accelerate particles along
field lines by centrifugal action (Emmering, Blandford \&
Shlosman 1992, Campbell 1999).  Bottorf et al. (1997) begin with
molecular material that is thrown up from the disk and becomes
exposed to the ionizing continuum, so creating broad emission
lines as the clouds are accelerated along the field lines. In
these models, as in the Murray et al. model, the range of radii
from which material exits the disk is large.

%%%%%%%%%%%%%%%%%%%%%%%%%%%%%%%%%%%%%%%%%%%
\subsection{Instabilities and Acceleration}
\label{instabilities}

Several instabilities are known that might create a wind from
special radii of an accretion disk.  The Lightman-Eardley
radiation pressure instability zone in the accretion disk
produces a wind originating within a critical instability radius,
and the velocities at this radius are comparable to the BELs
(Nicastro 2000).  The stabilizing effects of a corona may
restrict the range of radii in this model.  External radiation of
the disk can also create a wind from a restricted range of radii
at about the correct radius (Begelman, McKee, and Shields 1983;
Kurpiewski, Kurasczkiewicz and Czerny 1997). 

Nicastro (2000) also uses an `instability strip' in a disk, in
his case to explain the range of BEL widths.  The radii at which
the instability operates changes with {\em \.{m}}, the accretion
rate relative to the critical Eddington rate. Varying {\em \.{m}}
can explain the full range of BEL widths from
$\sim$1000~km~s$^{-1}$ for Narrow Line Seyfert~1s at high {\em
\.{m}}$\sim$1, to the broadest lines at $\sim$20,000~km~s$^{-1}$,
lines, which occur the lowest {\em \.{m}} consistent with the
disk instability lying outside the minimum stable orbit and so
still able to operate. LINERS may have {\em \.{m}} below this
value so that no BELR can form. Changing the radii having the
instability will change the opening angle of the flow (\S
\ref{BELR}). Some luminosity dependent changes (\S \ref{baldwin})
may then reflect a more fundamental change in \.{m}.
Coupling Nicastro's model with
the model presented here seems to explain most of the emission
and absorption line phenomena in AGN and quasars.

%%%%%%%%%%%%%%%%%%%%%%%%%%%%%%%
\subsection{Radio Loud Quasars}
\label{radio}

BALs have been found, until recently, only in radio-quiet quasars
(Stocke et al.~1992).  This suggested quite strongly that the
formation of BAL structures was intimately linked with a quasar's
radio properties.  No such link is implied in the proposed
structure.  How can we reconcile this disjunction?

\begin{figure*}[h]
\label{radioloud}
\centerline{
\psfig{figure=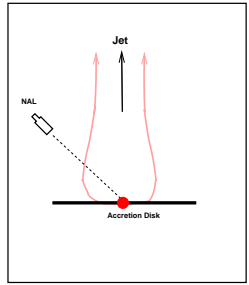,height=4in}
}
\caption{Geometry of a recollimated outflow in radio-loud
quasars.} 
\end{figure*}

There are (at least) three options:

\begin{enumerate}

\item \underline{Observational bias}: Goodrich (1997a) has shown
that modest amounts of dust can reduce the observed fraction of
BALs by a factor $\sim 3$.  So if radio-loud quasars are
preferentially dustier than radio-quiets they would be relatively
BAL-free.  Such quasars may be re-classified as `radio galaxies'
and so dropped from the samples (e.g. 3C22, Economou et al.
1995).  A second effect may be `beaming bias'.  Even a low
frequency selected radio sample contains many quasars in which a
beamed continuum dominates.  These are seen pole-on and, in the
proposed structure, would show no absorption.  The FIRST survey
has now found a substantial fraction of BALS (Becker et al. 1997,
1999, Gregg et al. 2000) and may be redressing these biases.

\item \underline{Jets destroy the BAL outflow}: The presence of a
relativistic jet will change the environment local to the flow.
Cosmic rays or radiation may overionize the flow, reducing the
radiation force. Alternatively the jet radiation or particle
pressure may disrupt the outer part of the flow with a non-radial
forces. Finally the harder continuum of a radio-loud quasar
(Elvis et al. 1994) may be less effective at accelerating the
outlfow.

\item \underline{The outflow recollimates to form the jet}: The
most elegant solution is that the outflow re-collimates to form
the jet. Jets are collimated and accelerated by means which are
much debated (e.g. Cellotti and Blandford 2000).  Even the scale
on which these effects occur is not well determined.  It is
possible then that, e.g. magnetic fields, re-collimate the
outflow near the point at which it would otherwise be accelerated
radially to BAL velocities, and instead accelerates the flow
toward the poles (figure~\ref{radio}) forming the jet. Kuncic (1999) and
Vestergaard, Wilkes \& Barthel (2000) make similar proposals.
Junor et al. (1999) see a similar geometry in VLBI observations
of M87: a 60$^{\circ}$ cone out to 100 Schwarzchild radii, where
the jet then becomes collimated.  If this suggestion is correct,
then all radio-loud quasars (apart from those dominated by a
beamed continuum) will show NALs (figure~\ref{radioloud}).
\end{enumerate}

%%%%%%%%%%%%%
%%%%%%%%%%%%

%%%%%%%%%%%%%%%%%%%%%%%%%%%%%%%%%%%%%%%%%%%%%%%%%%%%%%%%%%%%%%%%
\section{Conclusion}
\label{conclusion}

The structure proposed here is clearly ambitious. Yet, if we
believe that quasars are a solvable problem, some coherent
structure must be present. The present proposal, while in strong
need of elaboration, draws together previously disparate areas of
quasar research into a single simple scheme: the high and low
ionization parts of the broad emission line region (BELR), the
broad and narrow absorption line (BAL, NAL) regions, and the five
Compton thick scattering regions can all be combined into the
single funnel-shaped outflow. On large scales this outflow could
produce the bi-conical narrow emission line regions and, with
small geometric changes, several luminosity dependent effects can
be understood.  All of these features come about simply by
requiring a geometry and kinematics constructed only to explain
the two types of absorption lines.  This unification gives the
model a certain appeal.

%\notetoeditor{hi there}

%%%%%%%
\section*{Acknowledgments}

This work has been made possible by the many fascinating and
informative discussions I have had with my colleagues and friends
at CfA and elsewhere, in particular with Ski Antonucci, Jill
Bechtold, Nancy Brickhouse, Massimo Cappi, Bo\.{z}ena Czerny,
Giuseppina (Pepi) Fabbiano, Fabrizio Fiore, Margarita Karovska,
Andy Lawrence, Smita Mathur, Jonathan McDowell, Norm Murray,
Fabrizio Nicastro, Brad Peterson, Richard Pogge, Aneta
Siemiginowska, and Meg Urry. The 3-D renderings were kindly
produced by Antoine Visonneau of the Center for Design
Informatics at the Harvard School of Design using the
3D~StudioMax software package. This work was supported in part by
NASA contract NAS8-39073 (Chandra X-ray Center).

\newpage
%%%%%%%%%%%%%%%%%%%%%%%%%%%%%%%%%%%%%%%%%%%%%%%%%%%%%%%%%%%%%%
\section{References}

\refer{
Almaini O., Boyle B. J., Griffiths R. E., Shanks T., Stewart
G. C., \& Georgantopoulos I., 1995, {\it MNRAS} {\bf 277}, L31.}
\refer{
Antonucci R., \& Miller J.S., 1985, {\it Ap.J.}  {\bf 621}.}
\refer{
Baldwin J.A., 1977, {\it Ap.J.}, {\bf 214} 679.}
\refer{
Baldwin J.A., 1997, in ``Emission Lines in Active Galaxies: New
Methods \& Techniques'' (IAU Colloquium 159), eds. B.M. Peterson,
F.-Z. Chang \& A.S. Wilson, {\it PAS Conf. Proc.}, {\bf 113} 85.}
\refer{
Becker R.H., Gregg M.D., Hook I.M., McMahon R.G., White R.I., \&
Helfand D.J., 1997, {\it Ap.J.}, {\bf 479}, L93.}
\refer{
Becker R.H, White R.L., Gregg M.D., Brotherton M.,
Laurent-Muchleisen S., \& Arav N., 2000, {\it Ap.J.}, submitted.}
\refer{
Begelman M.C., Mc~Kee C.F. \& Shields G.A., 1983, {\it Ap.J.}, {\bf
271} 70.}
\refer{
Berriman G., Schmidt G.D., West S.C., \& Stockman H.S. 1990, {\it
Ap.J.S.}, {\bf 74}, 869.}
\refer{
Beloborodov A.M. \& Poutanen J., 1999, {\it Ap.J.Letters}, {\bf 517},
L77.}
\refer{
Binette L., 1998, {\it MNRAS}, {\bf 294}, L47}
\refer{
Bottorf M.C., Korista K.T., Shlosman I. \& Blandford R.D., 1997,
{\it Ap.J.}, {\bf 479}, 200}
\refer{
Bromley B.C., Miller W.A. \& Pariev V.I., 1998, {\it Nature} {\bf
391}, 54.}
\refer{
Campbell C.G., 1999, {\it MNRAS}, {\bf 310}, 1175}
\refer{
Capetti, A. Macchetto, F.D., \& Lattanzi, M.G., 1997, {\it Ap.J.},
{\bf 476}, L67.}
\refer{
Capetti, A., Axon, D.J., Macchetto, F.D., Sparks, W.B., \&
Boksenberg, A. 1996, {\it Ap.J.}, {\bf 466}, 166.}
\refer{
Cassidy I. \& Raine D.J., 1993, {\it MNRAS}, {\bf 260} 385.}
\refer{
Cassidy I. \& Raine D.J., 1996, {\it A\&A} {\bf 310}, 44.}
\refer{
Cellotti A. and Blandford R.D. 2000, in ``Black Holes in Binaries and
Galactic Nuclei'', eds.~L. Kaper, E.P.J. van den Heuvel, P.A. woodt
[springer-Verleg].}
\refer{
Chiang J., Reynolds C.S., Blaes O.M., Nowak M.A., Murray N.,
Madejski G., Marshall H.L. \& Magdziarz P., 2000, {\it Ap.J.},
{\bf 528}, 292}
\refer{
Clavel J. Wamsteker W. \& Glass I.S., 1989, {\it Ap.J.}, {\bf
337}, 236}
\refer{
Cohen M.H., Ogle P.M., Tran H.D., Vermuelen R.C., Miller J.S.,
Goodrich R.W. \& Martel A.R., 1995, {\it ApJ} {\bf 448}, L77.}
\refer{
Collin-Souffrin S., Dyson J.E., McDowell J.C. \& Perry J.J., 1988,
{\it MNRAS} {\bf 232}, 539.}
\refer{
Comastri A., Setti G., Zamorani G. \& Hasinger G., 1995, {\it A\& A}
{\bf 296}, 1.}
\refer{
Costatini E., et al., 2000, {\it ApJ}, submitted}
\refer{
Crenshaw D.M. \& Kraemer S.B., 2000, {\it Ap.J.Letters}, {\bf 532}, L101.}
\refer{
Crenshaw D.M., Kraemer S.B., Boggess A., Marran S.P., Mushotzky,
R.F., \& Wu C.-C., 1999, {\it Ap.J.}, {\bf 516}, 750.}
\refer{
Crenshaw D.M. and Kraemer S.B. 1999, {\it Ap.J.}, {\bf 521}, 572.}
\refer{
deKool M., 1997, ``Mass Ejection from AGN'', eds. N. Arav, I. Shlosman
\& R.J. Weymann, ASP Conf. Series, 128, 233.}
\refer{Done C., Madejski G.M. \& \.{Z}ycki P.T., 2000, {\tt
astro-ph/0002023}
}
\refer{
Economou F., Lawrence A., Ward M.J. \& Blanco P.R., 1995, {\it
MNRAS}, {\bf 272}, L5}
\refer{
Edelson R.A., Malkan M.A., Rieke G.F., 1987, {\it Ap.J.} {\bf 321},
233.}
\refer{
Elvis M., et al. 1994, {\it Ap.J.S.}, {\bf 95}, 1.}
\refer{
Emmering R.T., Blandford R.D. \& Shlosman I., 1992, {\it Ap.J.},
{\bf 385}, 460}
\refer{
Espey B.R. \& Andreadis S.J., 1999, ``Quasars and Cosmology'', ASP
Conf. Series, {\bf 162}, 351.}
\refer{
Fabian A.C., et al. 1994, {\it PASJ}, {\bf 46}, L59.}
\refer{
Fabian A.C., et al. 1995, {\it MNRAS} {\bf 277} L11.}
\refer{
Feldmeier A., Norman C., Pauldrach A., Owocki S., Puls J. \& Kaper L.,
1997, ``Mass Ejection from AGN', eds. N. Arav, I. Shlosman \&
R.J. Weymann, ASP Conf. Series, {\bf 128}, 258.}
\refer{
Ferland G.J., Korista K.T. \& Peterson B.M., 1990, {\it Ap.J.} {\bf
363}, L21.}
\refer{
Ferland G.J., Peterson B.M., Horne K., Welsh W.F. \& Nahar S.N., 1992,
{\it Ap.J.} {\bf 387}, 95.}
\refer{
Gallagher S.C., Brandt W.N., Sambruna R.M., Mathur S. \& Yamasaki,
1999, {\it Ap.J.}, in press.}
\refer{
Georgantopoulos I., Papadakis I., Warwick R.S., Smith D.A., Stewart
G.C. \& Griffiths R.G., 1999, {\it MNRAS} submitted ({\tt
astro-ph/9903083}).}
\refer{
George I.M. \& Fabian A.C., 1991, {\it MNRAS} {\bf 249}, 352.}
\refer{
George I.M., Turner T.J. \& Netzer H., 1995, {\it Ap.J.L.} {\bf 438},
L67.}
\refer{
Goad R.W. \& Koratkar A.P.,
1998, {\it Ap.J.}, {\bf 495} 718.}
\refer{
Goad R.W., Koratkar A.P., Axon A.J., Korista K.T. \& O'Brien P.T.,
1999, {\it Ap.J.(Letters)}, {\bf 512} L95.}
\refer{
Goodrich R.W. \& Miller J.S., 1994, {\it Ap.J.} {\bf 434}, 82.}
\refer{
Goodrich R.W. \& Miller J.S., 1995, {\it Ap.JL} {\bf 448}, L73.}
\refer{
Goodrich R.W., 1997a, ``Mass Ejection from AGN'', edS. N. Arav,
I. Shlosman \& R.J. Weymann, ASP Conf. Series, {\bf 128}, 94.}
\refer{
Goodrich R.W., 1997b, {\it Ap.J.}, {\bf 474}, 606.}
\refer{
Gregg M., et al. 2000, in preparation.}
\refer{
Habing H.J., 1996, {\it A \& A Rev.}, {\bf 7}, 97}
\refer{
Halpern J.P. \& Moran E., 1998, {\it Ap.J.}, {\bf 494}, 194.}
\refer{
Hamann F., Korista K.T. \& Morris S.L., 1993, {\it Ap.J.} {\bf 415},
541.}
\refer{
Hamann F., Shields J.C., Ferland G.J. \& Korista K., 1995a, {\it
Ap.J.} {\bf454}, 688.}
\refer{
Hamann F., Zuo L \& Tytler D., 1995b,  {\it Ap.J.L} {\bf444}, L69.}
\refer{
Hamman F., 1998, {\it Ap.J.}, {\bf 500}, 798.}
\refer{
Ho L.C., 1998, in `Observational Evidence for Black Holes in the
Universe', ed. S.K. Chakrabati [Dordrecht:Kluwer], p.   }
\refer{
Huchra J.P. \& Burg R., 1992, {\it Ap.J.} {\bf 393}, 90.}
\refer{
Impey C.D., Malkan M.A., Webb W. \& Petry C.E., 1995, {\it Ap.J} {\bf
440}, 80.}
\refer{
Iwasawa K. \& Taniguchi Y., 1993, {\it Ap.J.L} {\bf 413}, L15.}
\refer{
Iwasawa K., Fabian A.C., Young A.J., Inoue, H., \& Matsumoto C., 1999,
{\it MNRAS}, {\bf 306}, L19.}
\refer{
Junor, W., Biretta, J.A., \& Livio, M. 1999, {\it Nature}, {\bf 401},
891.}
\refer{
Kaastra J.S., Roos N. \& Mewe R., 1995,  {\it A\& A}, {\bf 300}, 25.}
\refer{
Kay L.E., 1994, {\it Ap.J.} {\bf 430}, 196.}
\refer{
Komossa S., 1999, in ``Structure and Kinematics of Quasar Broad Line
Regions'', ASP Conference Series, 175. eds. C. M. Gaskell,
W.N. Brandt, M. Dietrich, D. Dultzin-Hacyan \& M. Eracleous, p.365}
\refer{
Koratkar A., Antonucci R.R.J., Goodrich R.W., Bushouse H. \& Kinney
A.L., 1995, {\it Ap.J.} {\bf 450}, 501.}
\refer{
Korista, K.T., et al.1995, {\it Ap.J.S.}, {\bf 97}, 285.}
\refer{
Krolik J.H. \& Kriss G.A., 1995, {\it Ap.J.} {\bf 447}, 512.}
\refer{
Krolik J.H., 1999, ``Active Galactic Nuclei'' [Princeton: Princeton
University Press].}
\refer{
Krolik J.H., McKee C. \& Tarter B., 1981, {\it Ap.J.} {\bf 249} 422.}
\refer{
Kuncic Z. 1999, {\it PASP}, {\bf 111}, 954.}
\refer{
Kurpiewski A., Kurasczkiewicz J. \& Czerny B., 1997, {\it MNRAS} {\bf
285}, 725.}
\refer{
Lawrence A. and Elvis M., 1982,  {\it Ap.J.} {\bf 256}, 410.}
%\refer{
%Lee J.C., Fabian A.C., Brandt W.N., Reynolds C.S. \& Iwasawa K., 1999,
%{\it MNRAS}, in press.}
\refer{
Lee L.W. \& Turnshek D.A., 1995, {\it Ap.J.(Letters)}, {\bf 453}, L61}
\refer{
Lightman A.P. \& White T.R., 1988, {\it Ap.J.} {\bf 335}, 57.}
\refer{Madejski G.M., \.{Z}ycki P., Done C., Valinin A., Blanco
P., Rothschild R. \& Turek B., 2000, {\it Ap.J.Letters} in
press. {\tt astro-ph/0002063}
}
\refer{
Maiolino R. \& Rieke G.H., 1995, {\it Ap.J.} {\bf 454}, 95.}
\refer{
Maoz D., Edelson R. \& Nandra K., 1999, {\tt astro-ph/9910023}}
\refer{
Malkan  M.A., Gorjian V., Tam R., 1998, {\it Ap.J.(Supp.)}, {\bf 117},
25.}
\refer{
Marshall H.L., et al. 1997, {\it Ap.J.}, {\bf 479}, 222.}
\refer{
Mathur S., Elvis M. \& Singh K.P., 1996, {\it Ap.J.} {\bf 455}, L9.}
\refer{
Mathur S., Elvis M. \& Wilkes B.J., 1995, {\it Ap.J.} {\bf 452}, 230.}
\refer{
Mathur S., Elvis M. \& Wilkes B.J., 1999, {\it Ap.J.}, {\bf 519},
605.}
\refer{
Mathur S., Wilkes B.J. \& Elvis M., 1998, {\it Ap.J.(Letters)},
{\bf 503}, L23.}
\refer{
Mathur S., et al., 2000, {\it Ap.J.(Letters)}, in press.}
\refer{
Matt G., Fabian A.C. \& Ross R., 1991, {\it MNRAS} {\bf 262}, 179.}
\refer{
Matthews W.G., 1974, {\it Ap.J.} {\bf 189}, 23.}
\refer{
Matthews W.G., 1986, {\it Ap.J.} {\bf 305}, 187.}
\refer{
McHardy I. et al. 1995, {\it MNRAS} {\bf 273}, 549.}
\refer{
Miller J.S. \& Goodrich R.W., 1990, {\it Ap.J.} {\bf 355}, 456.}
\refer{
Miller J.S., Goodrich R.W.  \& Mathews W., 1991, {\it Ap.J.} {\bf
378}, 47.}
\refer{
Murray N. \& Chiang J., 1995, {\it Ap.J.L} {\bf 454}, L105.}
\refer{
Murray N. \& Chiang J., 1998, {\it Ap.J.}, {\bf 494}, 125.}
\refer{
Murray N., Chiang J., Grossman S.A. \& Voit G.M., 1995 {\it Ap.J.}
{\bf 451}, 498.}
\refer{
Nandra K., George I.M., Mushotzky R.F., Turner T.J., \& Yaqoob T.,
1997a, {\it Ap.J.}, {\bf 477} 602.}
\refer{
Nandra K., George I.M., Mushotzky R.F., Turner T.J., Yaqoob T., 1997b,
{\it Ap.J.}, {\bf 488} L91.}
\refer{
Netzer H., 1993, {\it Ap.J.} {\bf 473}, 781.}
\refer{
Nicastro F., 2000, {\it Ap.J.Letters}, 530, L65}
\refer{
Nicastro F., Fiore F., Perola G.C. \& Elvis M., 1999, {\it Ap.J.}
{\bf 512}, 184.}
\refer{
Nicastro F., et al, 2000, {\it Ap.J.}, in press. {\tt astro-ph/001201}}
\refer{
Ogle P.M., 1997, ``Mass Ejection from AGN'', eds. N. Arav, I. Shlosman
\& R.J. Weymann, ASP Conf. Series, {\bf 128,} 78.}
\refer{
Ogle P.M., 1998, PhD thesis, California Institute of Technology.}
\refer{
Osmer P.S., \& Shields J.C., 1999, ``Quasars and Cosmology'', ASP
Conf. Series, {\bf 162}, 235.}
\refer{
Osterbrock D.E., 1989, ``Astrophysics of Gaseous Nebulae and Active
Galactic Nuclei'' [Mill Valley: Univ. Science Books].}
\refer{
Peterson B.M., 1993, {\it PASP}, {\bf 105}, 247}
\refer{
Peterson B.M., 1997, `An Introduction to Active Galactic Nuclei'
[Cambridge:CUP].}
\refer{
Peterson B.M. \& Wandel A., 1999, {\it Ap.J.(Letters)}, 521, L95}
\refer{
Peterson B.M. \& Wandel A., 2000, in preparation}
\refer{
Pier E.A \& Krolik J.H., 1992, {\it Ap.J.(Letters)} {\bf 399}, L23.}
\refer{
Piro L., Yamauchi M. \& Matsuoka M., 1990, {\it Ap.J.} {\bf 360},
L35.}
\refer{
Porquet D., Dumont A.-M., Collin S. \& Mouchet M., 1999, {\it
A\&A},{\bf 341} 58.}
\refer{
Pounds K.A., Nandra K., Stewart G.C., George I.M. \& Fabian, A., 1990,
{\it Nature}, {\bf 344}, 132.}
\refer{
Proga D., Stone J.M., \& Drew J.E. 1998, {\it MNRAS}, {\bf 296}, L6.}
\refer{
Proga D., Stone J.M., \& Drew J.E. 1999, {\it MNRAS}, {\bf 310},
476.}
\refer{
Proga D., Stone J.M., \& Kallman T.R. 2000, {\tt
astro-ph/0005315}
}
\refer{
Reynolds C.S., 1997, {\it MNRAS} {\bf 286}, 513.}
\refer{
Rudge C.M. \& Raine D.J., 1998, {\it MNRAS}{\bf 297}, L1.}
\refer{
Rush B., Malkan M.A. \& Spinoglio L., 1993, {\it Ap.J.S} {\bf 89}, 1.}
\refer{
Salpeter E.E., 1974, {\it Ap.J.}, {\bf 193}, 585}
\refer{
Schild R.E., 1996, {\it Ap.J.}, {\bf 464}, 125}
\refer{
Schinnerer E., Eckart A., Tacconi L.J., Genzel R. \& Downes D.,
ApJ, in press ({\tt astro-ph/9911488}) }
\refer{
Schmidt G.D. \& Hines D.C., 1999, {\it Ap.J.}, {\bf 512}, 125.}
\refer{
Shields J.C., 1994, ``Reverberation mapping of the broad-line region
in active galactic Nuclei'', eds. Goodhalekar, P.M. Horne, K. and
Petterson, B.M., ASP Conference Servies 69, 293.}
\refer{
Shields J.C., Ferland G.J. \& Peterson B.M., 1995, {\it Ap.J.} {\bf
441}, 507.}
\refer{
Shull M.J. \& Sachs E.R., 1993, {\it Ap.J.} {\bf 416}, 536.}
\refer{
Simcoe J.A., McLeod K.K., Schachter, J. \& Elvis M., 1998, {\it
Ap.J.}, {\bf 489}, 615.}
\refer{
Stocke, J.J., Morris. S.L., Weymann, R.J. \& Foltz, C.B., 1992, {\it
Ap.J.}, {\bf 396}, 487.}
\refer{
Tadhunter C. \& Tsvetanov Z., 1989, {\it Nature} {\bf 341}, 422.}
\refer{
Tanaka Y. et al. 1995, {\it Nature}, {\bf 375}, 649.}
\refer{
Telfer R.C., Kriss G.A., Zheng W., Davidsen A.F. \& Green R.F., 1998,
{\it Ap.J.}, {\bf 509}, 132.}
\refer{
Turner T.J., Nandra K., George I.M., Fabian A.C. \& Pounds K.A., 1993,
{\it Ap.J.} {\bf 419},127.}
\refer{
Turnshek D.A., 1988 ``QSO Absorption Lines: Probing the Universe'',
eds. J.C. Blades, D.A. Turnshek, C.A. Norman [Cambridge:CUP], p. 17.}
\refer{
Turnshek D.A., Kopko M., Monier E., Noll D., Espey B.R. \& Weymann
R.J., 1996, {\it Ap.J.} {\bf 463}, 110.}
\refer{
Vestergaard M., 2000, {\it PhD Thesis}, Niels Bohr Institute for
Astronomy, Physics \& Geophysics, Copenhagen University}
\refer{
Vignali et al., 2000, {\it ApJ}, submitted}
\refer{
Vestergaard M., Wilkes B.J. \& Barthel P., 2000, {\it
Ap.J.(Letters)}, submitted }
\refer{
Wang J.X., Zhou Y.Y., Xu H.G., \& Wang T.G., 1999, {\it
Ap.J.(Letters)}, {\bf 516}, L65.}
%\refer{
%Weymann R.J., 1997, ``Mass Ejection from AGN'', eds. N. Arav,
%I. Shlosman \& R.J. Weymann, ASP Conf. Series, {\bf 128}, 3.}
\refer{
Weymann R.J., Morris S.L., Foltz C.B. \& Hewett P.C., 1991 {\it Ap.J.}
{\bf 373}, 23.}
\refer{
Wilkes B.J., 1984, {\it MNRAS} {\bf 207}, 73.}
\refer{
Williams R.J.R., 2000, {\it MNRAS}, submitted}
\refer{
Yaqoob T., Serlemitsos P.J., Turner T.J., George I.M. \& Nandra K.,
1996, {\it Ap.J.} {\bf 470}, L27.}
\refer{
Young S., Corbett E.A., Giannuzzo M.E., Hough J.H., Robinson A.,
Bailey J.A. \& Axon D.J., 1999, {\it MNRAS}, {\bf 303}, 227.}
\refer{
\.{Z}ycki P.T. \& Czerny B., 1994, {\it MNRAS} {\bf 266} 653.}

%%%%%%%%%%%%%%
\end{document}